\documentclass[twocolumn,floatfix,aps,showpacs,superscriptaddress]{revtex4} 

\usepackage{amsmath} 
\usepackage{graphicx}% Include figure files
\usepackage{dcolumn}% Align table columns on decimal point
\usepackage{bm}% bold math

\usepackage{epsfig}
\newcommand{\de}{\textrm{d}} 
\newcommand{\ee}{\textrm{e}} 
\newcommand{\ie}{\textrm{i}} 
\newcommand{\frc}[2]{\mbox{$\frac{#1}{#2}$}} 
\newcommand{\nuc}[2]{$^{#1}${#2}} 
\begin{document}
\bibliographystyle{apsrev}

\title{Continuum RPA study of the incoherent $\mu^- - e^-$ 
conversion rate \\ and its spurious $1^-$ admixture } 

\author{P. Papakonstantinou} 
\email[Email:]{panagiota.papakonstantinou@physik.tu-darmstadt.de} 
\affiliation{Institut f\"ur Kernphysik, Technische Universit\"at Darmstadt, D-64289 Darmstadt, Germany} 

\author{T.S. Kosmas} 
\affiliation{Department of Physics, University of Ioannina, GR-45110 Ioannina, Greece} 

\author{J. Wambach} 
\affiliation{Institut f\"ur Kernphysik, Technische Universit\"at Darmstadt, D-64289 Darmstadt, Germany} 

\author{A. Faessler} 
\affiliation{Institut f\"ur Theoretische Physik, Universit\"at T\"ubingen, D-72076 T\"ubingen, Germany}

\date{\today}

\begin{abstract}
The incoherent transition strength of the exotic $\mu^- - e^-$ conversion in
the $^{208}$Pb nucleus is investigated by utilizing the Continuum RPA method,
appropriate for the evaluation of the rate that 
goes to the continuum of the nuclear
spectrum. We find that the contribution of resonances 
lying high in the continuum is not negligible. 
Special attention is paid  
to the detailed study of the pronounced $1^-$
contribution which, according to previous calculations, dominates the overall incoherent rate 
in about all the nuclear 
targets. 
The spurious center of mass admixture to the partial rate originating from
the $1^-$ excitations is explored, 
and its elimination is performed by correcting 
properly the dipole operators. 
The results found this way show that the greatest 
portion of the total $1^-$ contribution 
to the incoherent rate is spurious. 
\end{abstract}


%\noindent
%KEYWORDS: Lepton flavor violation, muon-electron conversion, 
%Continuum RPA calculations, Spurious center of mass motion.

\maketitle

%
%%%%%%%%%%%%%%%%%%%%%%%%%%%%%%%%%%%%%%%%%%%%%%%%%%%%%%%%%%%%%%%%%%%%%%

\section{Introduction} 

The exotic neutrinoless conversion of a bound muon to an electron, 
\begin{equation}
\mu^-_b + (A,Z) \to e^- + (A,Z)^{\ast},
\label{mue}
\end{equation}
is an interesting lepton flavor violating process~\cite{Schaaf,Molzon,Kuno,tsk01}. 
One of its basic characteristics 
is the possibility of the coherent channel, i.e.  the ground state to ground
state transition \cite{KFV97,Okada}. 
Experimentally, only the ratio of the coherent rate divided by the total muon 
capture rate, which is the dominant branching ratio
exhausting a great part of 
the total $(\mu^-,e^-)$ rate %~\cite{KFV97}, %\cite{KFV97,Chiang}, 
could be measurable; 
by a judicious choice of the target nucleus, 
this channel could be free from 
the reaction induced background~\cite{Schaaf,Molzon,Kuno}. 

The incoherent $(\mu^-,e^-)$ rate is a less significant portion 
of the total rate and much harder to calculate, 
but its knowledge is important 
for determining the fraction of the coherent process 
to the total $(\mu^-,e^-)$ rate, 
which experimentally is also an interesting quantity. 
The theoretical calculation of the total rate requires 
reliable coherent and incoherent nuclear 
matrix elements~\cite{KV96,Suh-tsk,KFSV,SFK97}. 
Transitions of the reaction (\ref{mue}) have been 
previously studied by employing various methods, 
such as: (i) Closure
approximation within the shell model \cite{KV96} %\cite{KV89,KV89a} 
and the quasi-particle RPA (QRPA)  
%\cite{KVCF} 
for calculating the average contribution 
of the transitions to all excited states of the
target nucleus, (ii) a Fermi gas method utilizing a relativistic Lindhard 
function to compute the sum of all partial rates 
of the incoherent channel \cite{Chiang}, 
(iii) state-by-state calculations by using shell model \cite{Suh-tsk}, %,Suh-tskb}, 
and various QRPA versions \cite{tsk01,KFSV,SFK97}, 
to construct explicitly the final nuclear states. 

Within method (ii), the incoherent rate is calculated by 
integrating over 
a continuum of 
excited states of a local Fermi sea. 
Therefore, this method is not appropriate for individual 
calculations of each accessible channel, but  
it offers the advantage of taking into
consideration the part of the rate that goes to the continuum, 
which is not explicitly included in state-by-state calculations.
This is one of the reasons why the incoherent matrix elements obtained 
with shell model~\cite{Suh-tsk} and the various versions of RPA calculations~\cite{tsk01,KFSV,SFK97}, 
appear to be smaller 
compared to those of the method (ii). On the other hand, 
%\cite{KVCF,KFSV} compared to those of the method (ii). On the other hand, 
the common RPA and the various refinements 
of the QRPA 
%\cite{KVCF,KFSV},
offer a relatively simple and detailed state-by-state calculation of all the
individual low-lying excitations induced by the $\mu^- - e^-$ conversion operators
%\cite{KVCF,KFSV,tsk-01}.
\cite{KFSV,tsk01}.

An important conclusion of the state-by-state calculations is that 
the contribution of the $1^-$ states to the incoherent $(\mu^-,e^-)$ rate is 
very large (for most of the isotopes it is the maximum one) compared to that 
of the other multipolarities \cite{KFSV,SFK97}. The portion of the $1^-$ 
%of the other multipolarities \cite{KVCF,KFSV}. The portion of the $1^-$ 
contribution was found to be about $50\%$ for all mechanisms leading to the 
$\mu^- - e^-$ conversion. 
Therefore, it is essential to properly remove possible spurious 
contaminations when describing this process and other similar ones.  

The methods employed so far for the removal of the spurious center-of-mass 
(CM) admixture from the $1^-$ contribution \cite{Vehn,Pyatov} 
can be classified in two categories: 
(i) Those which remove from the contaminated Hamiltonian the spurious terms, 
i.e.,  those containing the CM position ($\bm{R}$) and the total momentum 
($\bm{P}$) operators and their couplings with the intrinsic Hamiltonian 
$H_{\textrm{int}}$ \cite{Vehn,Pyatov}. 
The diagonalization of $H_{\textrm{int}}$ obviously gives the 
real spectrum of the studied nucleus. 
In this way, the eigenstates of the system separate into the intrinsic 
nuclear spectrum and the pure CM excitation which can be omitted. 
Such methods, however, are tedious and not practical. 
(ii) Those which construct first a set of purified wave functions to be 
used for the diagonalization of the contaminated Hamiltonian. 
The orthonormalization, however, usually necessary in these methods, hinders 
their use. We should also mention that a recent method by Bes and Civitarese  
\cite{Civi-02}, which removes exactly the spurious
contaminations of the dipole operator, shows that the spurious portion 
is much bigger than previously thought and other methods give. 

In QRPA calculations, an effective elimination of spurious components 
from the $1^-$ states may be achieved by adjusting the parameters which 
scale the effective interaction, %G-matrix elements in the nuclear medium, 
so that the energy of the first 
$1^-$ state becomes equal to zero (equal to the purely spurious center
of mass eigenstate). In most of the cases, this requires unphysical values 
for the parameters which renormalize the individual particle-particle ($g_{pp}$) 
and particle-hole ($g_{ph}$) channels (usually $g_{pp} \approx 1.3-1.5$, 
$g_{ph} \approx 0.2-04$). 
Even %in self-consistent RPA models based on energy functionals - 
when the spurious state occurs very close to zero energy, spurious 
admixtures remaining at higher energies cannot be avoided completely. 

In Ref. \cite{SFK97} an approximate removal of the spurious $1^-$ components 
was performed by constructing the properly normalized purely spurious state 
$|S\rangle$ and evaluating its overlap with all $1^-$ states involved 
in the chosen model space. 
This showed that mostly the lowest lying $1^-$ state is affected by the 
translational invariance breaking caused by the use of empirical single 
particle energies and a truncated model space in RPA. 
This state was considered as fully spurious and the others were treated 
as real nuclear excitations. After removing the spurious contributions, 
a renormalization of the interaction was required for reproducing 
the energy spectrum of the nucleus with the use of realistic two-body forces. 
The above method is simple and easy to apply, 
but it is not as exact as that of Refs. \cite{Vehn,Civi-02}.

The purpose of the present work is to study in detail the incoherent 
rate of the $\mu^- - e^-$ conversion,  
by using the 
Continuum RPA method. 
We evaluate the contribution of high-lying continuum excitations. 
Such an explicit calculation has not yet been addressed. 
As an application, we study extensively the incoherent rate for the heavy nuclear target
$^{208}$Pb by using Skyrme interactions.
In order to eliminate spurious CM contaminations 
in the dominant $1^-$ channel, 
we obtain the corresponding energy distributions 
by using properly corrected dipole operators, which
induce the $\mu^- - e^-$ conversion intrinsic $1^-$ excitations. 
%The correction achieved is found quite large.

The paper is organized as follows. In Sec.~\ref{Defs} we describe 
briefly the $\mu^- -e^-$
conversion operators and the formalism of the 
Continuum RPA method. In Sec.~\ref{Res} we calculate the
strength distributions as functions of the excitation energy. The 
elimination of the spurious center of mass contaminations is also discussed. 
In Sec.~\ref{Sum} we summarize our main conclusions.

\section{Definitions and brief description of the method of calculation} 
\label{Defs} 
 
The inclusive $(\mu^-,e^-)$ rate is evaluated 
by summing the partial contribution of all final states $|f\rangle$. 
For spherical or nearly spherical nuclei, the vector contribution is given by~\cite{tsk01} 
\begin{equation} 
S_a=\sum_f\left(\frac{q_f}{m_{\mu}}\right)^2 |\langle f | 
O_a(\bm{q}_f) | 0 \rangle |^2 , 
\label{vec-con}  
\end{equation} 
where $O_a(\bm{q}_f)$ represents the $\mu^- - e^-$ 
vector-type transition operator
resulting in the context of a given mechanism mediated by a photon ($a=\gamma$),
a $W$-boson ($a=W$) or a $Z$-particle exchange ($a=Z$). 
Here $\bm{q}_f$, with magnitude  
$q_f=m_{\mu}-\epsilon_b-E_f$, 
is the momentum transferred to the nucleus. 
$E_f$ is the energy of the final state $|f\rangle$ with 
respect to the ground state $|0\rangle$, 
$\epsilon_b$ is the binding energy of the muon and $m_{\mu}$ its mass. 
The transition operators have the form 
\begin{equation} 
O_a(\bm{q}) =  \tilde{g}_Vf_V 
\sum_{j=1}^A 6c_a(\tau_{ j }) 
\ee^{-\ie \bm{q}\cdot\bm{r}_j} ,  
%\hspace{3mm} ; \hspace{3mm}  
\qquad 
c_a(\tau_{ j }) \equiv \frc{1}{2}+\frc{1}{6}\beta_a\tau_{ j } 
,  
\label{Eiqrop}  
\end{equation} 
where $\tau_{ j }$ is the 3rd component of the $j$th particle's isospin. 
The parameter $f_V=1.0$ represents the vector static nucleon 
form factor and the normalization coefficient $\tilde{g}_V$ 
takes the value $1/6$ for the photonic case and $1/2$ for the 
non-photonic $W$ boson and SUSY $Z$ exchange \cite{KFV97}. 
The value of $\beta_a$ depends on the model assumed. 
(We have taken the relevant values from Ref.~\cite{Suh-tsk}.) 
Thus, protons (neutrons) contribute to 
a given process with a ``charge" whose 
value is determined by $c_{a}(1/2) = 1/2+\beta_a /6$ ($c_{a}(-1/2)=1/2-\beta_a /6$).  
In the photon and $Z$ case, the isoscalar and isovector components of 
the transition operator are (almost) equally important, 
whereas $O_W$ is predominantly isoscalar.

By assuming that the initial and final states are of definite spin
and parity, a multipole decomposition of the operators 
%of Eqs.~(\ref{vec-con}), (\ref{Eiqrop}) 
of Eq.~(\ref{Eiqrop}) 
into operators $T_{aLM}$ of orbital angular momentum rank $L$ 
can be carried out. For spherical nuclei we can assume, 
without loss of generality,  
$\hat{q}=\hat{z}$. Then, only terms with 
$M=0$ survive, for which we obtain 
\begin{eqnarray} 
T_{aL}({q}) \equiv 
T_{aL0}({q}) 
&=&  \tilde{g}_Vf_V 
\sqrt{4\pi (2L+1)}  
\nonumber \\ 
& & \times 
\sum_{j=1}^A 6c_a(\tau_j) 
j_L(qr_j) Y_{L0} (\hat{r}_j) 
 . 
\label{Ejqrop}  
\end{eqnarray} 
A phase factor $(-\textrm{i})^L$ has been omitted. 
The contribution of each multipolarity to 
the transition rate $S_a$ reads 
\begin{equation} 
S_{aL}=\sum_f \left( \frac{q_f}{m_{\mu}} \right)^2 
| \langle f | 
T_{aL}(q_f)|0\rangle |^2 . 
\end{equation} 

We now rewrite the rate $S_{aL}$ as the integral of 
a suitable distribution over excitation energy:  
\begin{equation} 
S_{aL}  \equiv \int\de E R_{aL}(E) 
\label{e:sint} 
\end{equation} 
with 
\begin{eqnarray} 
R_{aL}(E) &=&  
\sum_f 
\left( 1 - \frac{\epsilon_b+E_f}{m_{\mu}} \right)^2 
\nonumber \\ 
 & & \times  
| \langle f | 
T_{aL}( 
m_{\mu}-\epsilon_b-E_f  
)|0\rangle |^2 
\delta (E-E_f) 
\nonumber \\ 
&=& 
\left[ 
\left( 1-\frac{\epsilon_b}{m_{\mu}}\right)^2  
-\frac{2}{m_{\mu}^2}(m_{\mu}-\epsilon_b)E 
\right. 
\nonumber \\ 
&& 
\left. 
+\frac{1}{m_{\mu}^2} E^2  
\right] 
R'_{aL}(E).  
\end{eqnarray} 
In the above expressions, 
\begin{equation} 
R'_{aL}(E)= \sum_f 
| \langle f | 
T_{aL}( 
m_{\mu}-\epsilon_b-E_f   
)|0\rangle |^2 
\delta (E-E_f) 
\label{Espr}  
\end{equation}  
stands for the ``strength distribution" 
corresponding to the operator $T_{aL} (q)$, with  
$q = m_{\mu}-\epsilon_b-E$. 
The total rate of Eq.~(\ref{e:sint}) is subsequently written as 
\[ S_{aL} = 
\left( 1-\frac{\epsilon_b}{m_{\mu}}\right)^2 M_0  
-\frac{2}{m_{\mu}^2}(m_{\mu}-\epsilon_b) M_1  
+\frac{1}{m_{\mu}^2} M_2 ,  
\] 
where 
$ %begin{equation} 
M_k \equiv \int R'_{aL}(E) E^k \de E 
$ %end{equation} 
is the $k-$moment of $R'_{aL}(E)$.  
The final states $|f\rangle$, excited by the single-particle operator $T_{aL}$, 
are of particle-hole ($ph$) type. 
Then, the distribution $R'_{aL}(E)$, 
and from it $R_{aL}(E)$ and $S_{aL}$, 
can be calculated 
following the standard RPA method. 

We consider $ph$ excitations, built on top of the Hartree-Fock (HF) 
ground state of a closed-shell nucleus 
and subjected to the $ph$ 
residual interaction (HF+RPA method). In this work, 
the quantities introduced above 
are calculated 
using a self-consistent  
Skyrme-Hartree-Fock (SHF) plus Continuum-RPA (CRPA) model. 
The HF equations describing the ground state 
are derived variationally from the Skyrme energy functional and 
solved numerically using the code of P.-G.~Reinhard \cite{ReXX}. 
The CRPA  
is formulated 
in coordinate space, 
as described, e.g., in 
\cite{VaS1981a,RWH1988} 
%\cite{BeXX,BeT1975,VaS1981a,vGi1983,RWH1988,PP2004} 
%%\cite{VaS1981a,RWH1988,PP2004} 
and outlined below. 

The main ingredient of the model is the 
$ph$ Green function $G_L(E)$ in coordinate space.  
In particular, we are interested in the radial part  
$G_{Lij}(r\tau,r'\tau';E)$, which  
describes the 
propagation of a fluctuation (or $ph$ state) 
of multipolarity $L$ and energy $E$, 
excited by the operator $V_{Li}$  
at the point $r,\tau$ 
and decaying via the operator $V_{Lj}$ 
at the point $r'\tau'$ 
 - where 
$\tau$ or $\tau'$ 
corresponds to the isospin character (proton or neutron) 
of the fluctuation. $V_{Li(j)}$ stands for one of the 
rank-$L$ operators 
%$\{$ 
$Y_{L}$, 
$ [Y_L\otimes (\nabla^2+{\nabla '}^2) ]_{L}$, 
$ [Y_{L \pm 1}\otimes (\bm{\nabla} - \bm{\nabla '})]_{L}$,  
 $ [Y_{L \pm  1}\otimes (\bm{\nabla} + \bm{\nabla '})]_{L}  $ 
%$\}\otimes\{ 1, \hat{\tau} \}$ 
present in the Skyrme interaction. 
In practice, 
the radial coordinates are replaced by  
points on a discretized mesh. 
A radial step $\Delta r$ and a maximum value $r_{\max}$ 
(larger than the nuclear radius by a factor of about 3 or more, usually) 
are introduced. 
Then the Green function $G_{L}(E)$ 
can be represented as a super-matrix 
in coordinate space, isospin character and operator indices $i$. %s $V_{L,i}$. 

The operator $T_{aL}(q)$, Eq.~(\ref{Ejqrop}), is a multipole operator 
of the type $Y_L$ - let us label it as $V_{L1}$. 
Therefore, within our RPA model, 
the distribution 
$R'_{aL}(E)$ of Eq.~(\ref{Espr}) is given 
in terms of the RPA Green function $G_L^{\textrm{RPA}}(E)$ by 
\begin{equation} 
R'_{aL}(E) = \frac{\textrm{Im}}{\pi} {\textrm{Tr}}[T_{aL}^{\dagger}(q)
G^{\textrm{ RPA}}_{L11}(E)T_{aL}(q)]  
\end{equation} 
in a matrix notation. 
In practice, this reads 
\begin{eqnarray} 
R'_{aL}(E) &=& 144 \tilde{g}_V^2f_V^2   
(2L+1) 
\sum_{\tau,\tau'} c_a(\tau )c_a(\tau ')  
\nonumber \\ 
 &&\times 
{\textrm{Im}} 
\int j_L(qr)G_{L11}^{\textrm{RPA}}(r\tau,r'\tau';E) 
 \nonumber \\ && \times j_L(qr') \de r \de r' 
 , 
\end{eqnarray} 
$q=m_{\mu}-\varepsilon_b-E$. The integrations are to be understood as numerical ones, 
carried out by summing over 
the radial mesh points. 

The RPA Green function is given by 
the equation 
\begin{equation} 
G_L^{\textrm{RPA}}(E) = [ 1+G_L(E)^0V_{\textrm{res}}]^{-1} G_L^0(E) 
 , 
\label{Egrpa} 
\end{equation} 
which is solved as a matrix equation in 
coordinate space, 
isospin character 
and 
operators $V_{Li}$. 
The $ph$ {residual interaction} 
$V_{\textrm{res}}$ is zero-range, 
of the Skyrme type, 
derived self-consistently from the Skyrme-HF energy 
functional 
%\cite{vGi1983,PP2004,ShB1975,Tsa1978}.   
%%\cite{PP2004,ShB1975,Tsa1978}.   
\cite{ShB1975,Tsa1978}.   
In this work, spin-dependent terms and the Coulomb interaction are 
omitted from $V_{\textrm{res}}$. 
%%%%\item 
% 
% 
The radial %and isospin 
part of the unperturbed $ph$ Green function 
of multipolarity $L$ 
is formally given by: 
\begin{eqnarray} 
G_{Lij}^0  
(r\tau, r'\tau ';E) 
&=& \delta_{\tau\tau'}  
\sum_{ph} \left\{  
\frac{ 
\langle p | V_{Li} | h \rangle^{\ast}_{r\tau} 
\langle p | V_{Lj} | h \rangle_{r'\tau}     
	  }{ 
      \varepsilon_{ph} -E} 
\right.  
 \nonumber \\ && \pm 
\left. 
\frac{ 
\langle h | V_{Lj} | p \rangle^{\ast}_{r'\tau} 
\langle h | V_{Li} | p \rangle_{r\tau}        
	  }{ 
     \varepsilon_{ph} +E} 
     \right\}  .  
\label{Eghf} 
\end{eqnarray} 
The sign of the second term depends 
on the symmetry properties 
of the operators $V_{Li}$ and $V_{Lj}$  
under parity and time-reversal 
transformations. 
With $h$~% 
($
p 
$) 
we denote 
the quantum numbers of the   
 HF hole~(particle) state and  
 $\varepsilon_{ph}=\varepsilon_p-\varepsilon_h$ is the energy 
 of the unperturbed $ph$ excitation. 
A small but finite Im$E\equiv\Gamma /2$ ensures that bound transitions 
acquire a finite width. % \cite{BeXX}. 

The particle continuum is fully taken into account, 
as follows 
%\cite{BeXX,vGi1983,PP2004,ShB1975}: 
%%\cite{PP2004,ShB1975}: 
\cite{ShB1975}: 
The summation over the particle states $p$ in Eq.~(\ref{Eghf}) 
is replaced by the summation over all single-particle states $k$. 
The additional hole terms in the first term on the r.h.s 
will cancel the ones in the second term. Next, $\varepsilon_k$ 
is replaced by the single-particle Hamiltonian. Finally, the completeness 
of the $k$ states and the properties of the particle Green function 
are  used to replace the sum over $k$ by a closed expression. 
Therefore, the only truncation introduced is the one of the radial 
coordinate, $r\leq r_{\max}$. 
The latter is very well justified, 
because the amplitude of the radial wavefunctions of the hole states entering expression 
(\ref{Eghf}) vanishes at distances much larger than the nuclear radius.   

Results derived within this model for the nucleus \nuc{208}{Pb} are presented
and discussed in the next section. 
 
%%%%%%%%%%%%%%%%%%%%%%%%%%%%%%%%%%%%%%%%%%%%%%%%%%%%%%%%

\section{Results and Discussion} 
\label{Res} 

In the following we will present results obtained using 
%For the purposes of the present work we have employed 
the SkM*\cite{BQB82} parametrization of the Skyrme force. 
It describes satisfactorily  giant resonances of stable nuclei, 
and therefore it is suitable for the present study. 
We have verified that our conclusions do not change when 
the parametrization SGII \cite{VaS1981a,VaS1981b} is used.   
In order to test the sensitivity of our results on the 
interaction used, we have also employed MSk7 \cite{GTP2001}, 
which has a large effective mass, thereby shifting 
most excited states to lower energies compared to the 
more reliable SkM* and SGII. 

For the nucleus \nuc{208}{Pb}, the muon binding energy is
$\epsilon_b = 10.475$~MeV and the momentum transferred to the nucleus by the
outgoing $e^-$ of the $\mu^- - e^-$ conversion ranges from $q=0.482$~fm$^{-1}$, 
when the transition energy $E$ vanishes (namely in the coherent process), to zero, 
when $E$ reaches the maximum value, i.e. $E_{\max{}}=m_{\mu}-\epsilon_b=95.183$~MeV
(namely when all the available energy of the bound $\mu^-$ goes to a nuclear
excitation).  
The particle threshold energy $E_{\textrm{thr}}$ 
is 8.09~MeV %~(7.58~MeV)  
in the case of the SkM* force. %~(MSk7) force. 
We have obtained results for $L=0,1,\ldots,6$ and 
for natural parity, $(-1)^L$. 
The most important contributions to the 
incoherent transition rate are expected from 
$L<4$ \cite{SFK97}.  
In all cases %, for the finite Im$E$
%, except for dipole transitions, 
we have used $\Gamma = 0.2$~MeV and $r_{\max} = 17$~fm. 

\subsection{Incoherent transition-strength distributions vesrus excitation energy } 

\begin{figure} 
\includegraphics[width=0.8\columnwidth]{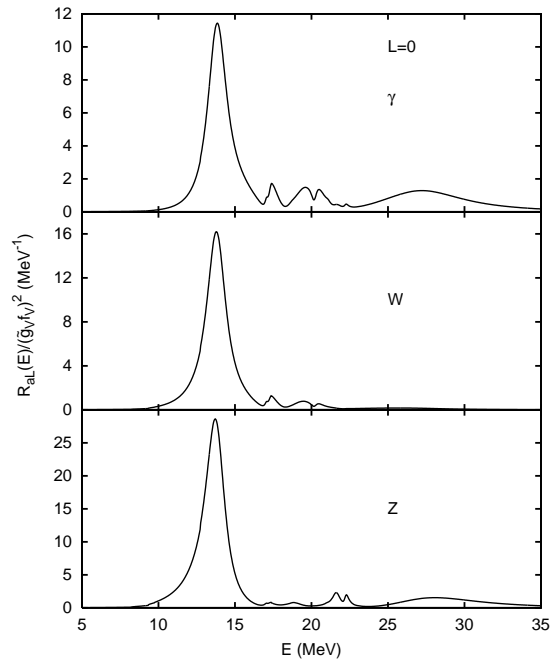} 
\caption{% 
\label{Fstot1a}  
The distribution $R_{aL}(E)$ in $^{208}$Pb for $L=0$. 
Skyrme parameterization SkM* has been used.}  
\end{figure}

\begin{figure} 
\includegraphics[width=0.8\columnwidth]{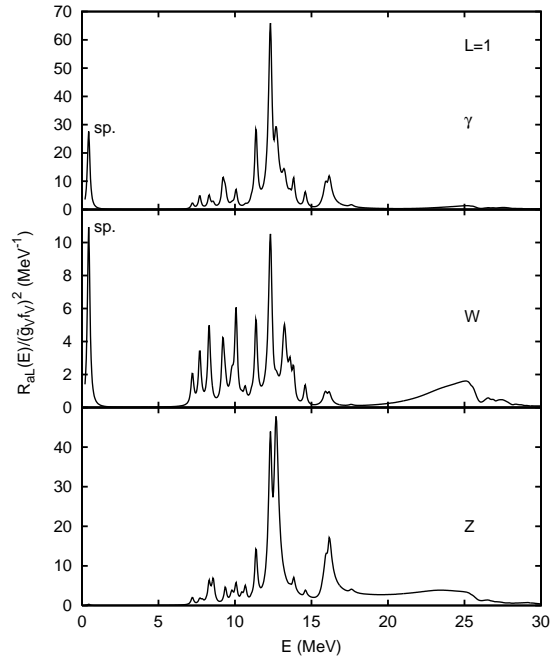} 
\caption{Same as Fig. \ref{Fstot1a}, for $L=1$. 
\label{Fstot1b}  
}  
\end{figure}

\begin{figure} 
\includegraphics[width=0.8\columnwidth]{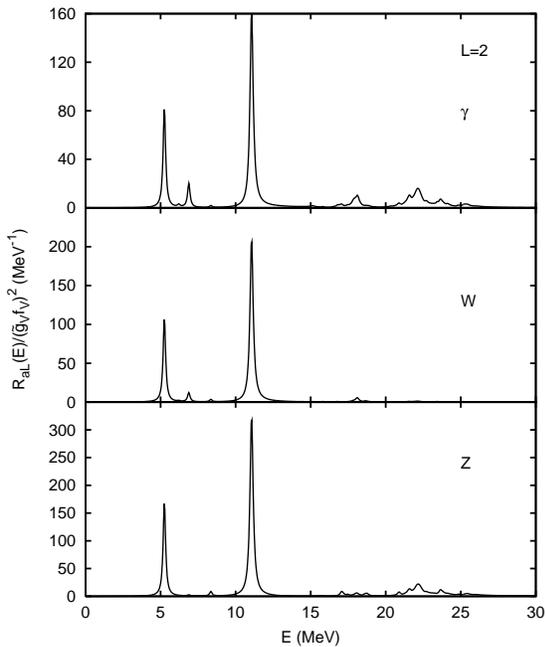} 
\caption{Same as Fig \ref{Fstot1a}, for $L=2$. 
\label{Fstot1c}  
}  
\end{figure}

\begin{figure} 
\includegraphics[width=0.8\columnwidth]{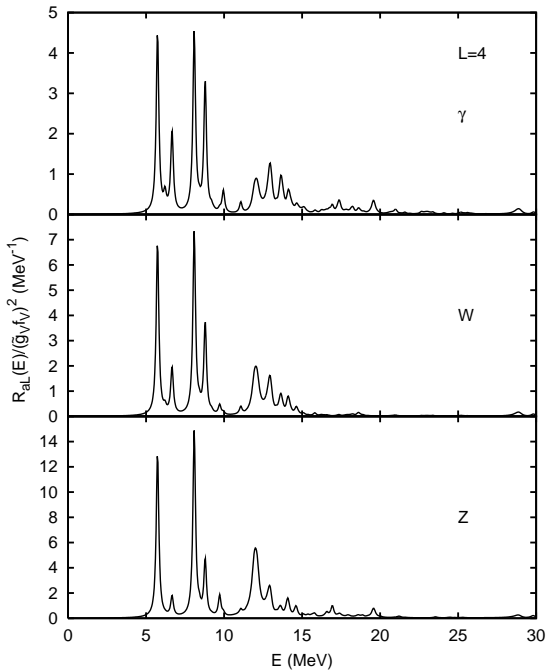} 
\caption{Same as Fig \ref{Fstot1a}, for $L=4$. 
\label{Fstot1e}  
}  
\end{figure}

In Figs.~\ref{Fstot1a}-\ref{Fstot1e}  
the distribution $R_{aL}(E)$ is plotted as a function of $E$, for $L=0,1,2,4$, 
and for natural-parity transitions. 
In the monopole case, $L=0$, 
the Isoscalar (IS) Giant Monopole Resonance (GMR)  is the main peak. 
For $\gamma$ and Z, there is considerable contribution coming
from higher energies (20-35~MeV), i.e., the isovector (IV) 
GMR region. 
For $L=1$, 
the IV Giant Dipole Resonance (GDR) 
corresponds to the strength 
clustered around $E \approx 12$~MeV. 
For $W$ and $Z$, important contribution seems to come 
from higher energies 
(above 20~MeV), in particular, the IS GDR. 
For $W$ exchange, the region below 10 MeV contributes 
significantly. In this region we find the oscillation 
of the neutron skin against the nuclear core 
(pygmy dipole resonance)~\cite{SIS90}. 
The transition density is isoscalar in the interior 
of the nucleus, while on the surface the proton contribution vanishes. 
In the quadrupole case, $L=2$, 
the IS Giant Quadrupole Resonance (GQR) is the second peak, 
close to 11~MeV. 
The collective 
low-lying peak is strong as well. 
There is some contribution from energies higher than 
15~MeV, i.e. from the IV GQR region, 
especially in the cases $\gamma$  and $Z$. 
For $L=3$ (not shown) the strength is mostly 
concentrated in the collective octupole state 
at low energy. 
For $L>3$, as shown in Fig.~\ref{Fstot1e} for $L=4$, 
the calculated strength is quite fragmented and 
most of it lies below 20~MeV. 

%%%%%%%%%%%%%%%%%%%%%%%%%%%%%%%%%%%%%%%%%%%%%%%%%%%%%%%%%%%%%%%%%%%%%%%%%%

%%%%%%%%%%%%%%%%%%%%%%%%%%%%%%%%%%%%%%%%%%%%%%%%%%%%%%%%%%%%%%%%%%%%%%%%%%

%%%%%%%%%%%%%%%%%%%%%%%%%%%%%%%%%%%%%%%%%%%%%%%%%%%%%%%%%%%%%%%%%%%%%%%%%%

%%%%%%%%%%%%%%%%%%%%%%%%%%%%%%%%%%%%%%%%%%%%%%%%%%%%%%%%%%%%%%%%%%%%%%%%%%

A finite momentum transfer $q_f$ can result in the 
excitation of overtones of giant resonances \cite{SKA03,PWP2004b}.  
With the exception of the IS GDR, such states lie typically above 
30~MeV, for \nuc{208}{Pb},  where  
the corresponding momentum transfer is less than 0.33~fm$^{-1}$. 
%the energy range plotted in Figs.~\ref{Fstot1}-\ref{Fstot4}. 
As a result, the possible contribution of such states was too small to be 
identified and visible on graphs.

The dipole results presented in Fig.~\ref{Fstot1b} were obtained with 
corrected dipole operators - see Sec.~\ref{Scmm} for more details 
regarding this special case. 
A small amount of spurious strength remains close to zero energy 
(labeled ``sp." on the figure), 
%when the force SkM* is used, 
but it is well separated from the 
rest of the distributions. 

\begin{figure} 
\includegraphics[width=0.8\columnwidth]{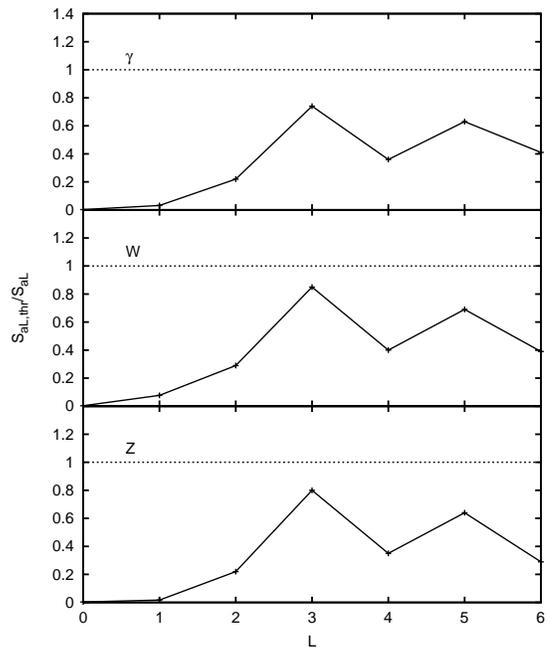} 
\caption{% 
\label{Fst20a}  
Fraction of the 
total strength $S_{aL}$, 
coming from states below the particle threshold 
vs. the multipolarity $L$. 
%Skyrme parameterizations SkM* and MSk7 have been used.  
Skyrme parameterization SkM* has been used.  
(Lines %connecting results obtained with the same Skyrme parameterization 
are drawn to guide the eye.) } 
\end{figure}

\begin{figure} 
\includegraphics[width=0.8\columnwidth]{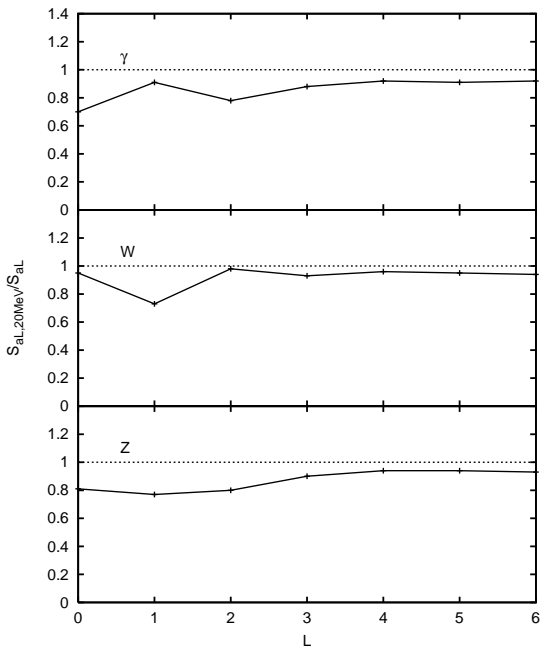} 
\caption{% 
\label{Fst20b}  
As in Fig.~\ref{Fst20a}, fraction of 
strength $S_{aL}$, 
coming from states 
below 20~MeV.}  
\end{figure}

In Fig.~\ref{Fst20a} we plot the fraction of the total strength $S_{aL}$ 
coming from states below the particle threshold 
($S_{aL,\textrm{thr}}$) 
and in Fig.~\ref{Fst20b} the fraction coming from states below 20~MeV 
($S_{aL,20\textrm{MeV}}$),  
vs. the multipolarity $L$. 
In the dipole case, the remaining strength 
of the spurious state 
%below 6~MeV (see Fig.~\ref{Fstot}) 
is not taken into account 
when evaluating these fractions.

%%%%%%%%%%%%%%%%%%%%%%%%%%%%%%%%%%%%%%%%%%%%%%%%%%%%%%%%%%%%%%%%%
%%%%%%%%%%%%%%%%%%%%%%%%%%%%%%%%%%%%%%%%%%%%%%%%%%%%%%%%%%%%%%%%%
%%%%%%%%%%%%%%%%%%%%%%%%%%%%%%%%%%%%%%%%%%%%%%%%%%%%%%%%%%%%%%%%%
%%%%%%%%%%%%%%%%%%%%%%%%%%%%%%%%%%%%%%%%%%%%%%%%%%%%%%%%%%%%%%%%%

We see that for low multipoles $L=0,1,2$ only a small portion of the
strength originates from energies below particle threshold.
The trend followed is similar for all mechanisms
and, as we have verified, 
independent of the interaction used. For even multipoles $L=2,4$
a big portion of the contribution is pushed to higher energies as
compared to the neighboring odd ones. Another interesting feature
is the fact that, for some multipoles ($L=0$ for photonic mechanism,
$L=1$ for $W$-boson exchange), a significant portion of the strength
comes from above $E=20$MeV.

We have also calculated the  fraction of the total strength $S_{aL}$, 
coming from states below 50~MeV, for $L$ up to 6. 
In all cases, the fraction is practically equal to unity. 
This means that the discretized versions of RPA and QRPA 
for the examined nucleus \nuc{208}{Pb} are safe to use if 
the energy cutoff is large enough to sufficiently account for transitions 
below this value. 
 
\subsection{Dipole strength and the spurious CM motion} 
\label{Scmm} 

It is well known \cite{Vehn,Pyatov} that the $1^-$ excitations contain 
admixtures of the spurious excitation of the center of mass (CM) of the nucleus
\begin{equation}
\bm{R} \, = \, \frac{1}{A} \sum_{j=1}^A \, \bm{r}_j
\label{cmop}
\end{equation}
corresponding to a situation in which the unexcited nucleus moves as a 
whole around the localized fictitious potential well. Normally, these 
spurious components are separated out by the RPA methods. However, the 
use of a truncated model space and non-self-consistent single particle 
energies in ordinary RPA and the other versions of QRPA destroys 
the translational invariance and inserts spurious
excitations into the spectrum. Thus, the spurious CM 
state is not completely separated from the real (intrinsic) nuclear 
excitations, and in addition its energy eigenvalue is not zero. 

In Continuum-RPA models with Skyrme interactions it has been possible 
to achieve a high degree of self-consistency, i.e. the same interaction 
is used for the HF calculation of ground state properties and for the 
residual interaction. In addition, no truncation is involved. However, due 
to the formulation of the model in coordinate space, it is 
common practice to exclude the Coulomb and spin-orbit contribution 
(at least) to the residual interaction. Therefore, self-consistency is 
violated and, even in cases where the spurious state appears very close 
to zero energy, some spurious strength may remain at higher energies. 

For electric dipole excitations, the problem is usually treated by using effective 
charges \cite{BM69}. 
Similarly, in the case of IS dipole excitations, effective operators 
are used \cite{VaS1981a,ASS03}, 
which minimize the spurious admixture in the strength 
distribution. 
The effect on the IS dipole excitations of \nuc{208}{Pb} was examined in detail in 
Ref.~\cite{HaS2002b}.  
Here we will present a similar prescription for the operators 
involved in $\mu^- - e^-$ conversion. 

We begin with a dipole excitation operator of the generic form 
\begin{equation} 
\Omega_1=\sum_{j=1}^{A}c(\tau_j)f(r_j)Y_{10}(\hat{r}_j) = 
         \sum_{j=1}^{A}c(\tau_j)\frac{f(r_j)}{r_j} \sqrt{\frac{3}{4\pi}}z_j 
\label{Eomg} 
. 
\end{equation} 
As far as intrinsic excitations are concerned, this operator is 
equivalent to 
the corresponding ``corrected" operator 
\begin{equation} 
\Omega_1^{\textrm{corr}} \equiv \Omega_1 - \tilde{\eta} R_z  
=\sum_{j=1}^{A}[c(\tau_j)f(r_j)-\eta r_j] Y_{10}(\hat{r}_j) . 
\label{Eomgc}  
\end{equation} 
where $R_z$ stands for the $z-$component of the 
CM space vector $\bm{R}$ of Eq. (\ref{cmop}) and 
$$
\eta = \frac{1}{A}\sqrt{\frac{4\pi}{3}} \ \tilde{\eta}.
$$ 
Our task is to determine the parameter $\eta$ 
so as to eliminate the spurious CM excitation. 
This can be achieved within the collective model, by 
imposing the translational-invariance condition on the transition density 
characterising the  
collective state 
induced by $\Omega_1^{\textrm{corr}}$ 
\cite{VaS1981a,BM69}. A condition on $\eta$ is thus obtained 
analytically.  
The result is ($c_p\equiv c(1/2), c_n\equiv c(-1/2)$): 
\begin{equation} 
\eta = \frac{1}{3} 
\left[ 
\frac{c_pZ}{A}\langle \frac{1}{r^2} \frac{\de}{\de r} [f(r)r^2] \rangle_p 
+\frac{c_nN}{A}\langle\frac{1}{r^2}\frac{\de}{\de r} [f(r)r^2] \rangle_n 
\right] , 
\label{Eetag} 
\end{equation} 
where the mean values are defined as 
\[ 
\langle g(r) \rangle_{p,n} 
= \frac{\int_0^{\infty} g(r) \rho_{p,n}(r) r^2 \de r }{ 
\int_0^{\infty} \rho_{p,n}(r) r^2 \de r } 
\] 
and $\rho_p(r)$, $\rho_n(r)$, are the proton, neutron, densities in the nuclear 
ground state. They are normalized so that $4\pi \int_0^\infty \rho_{p,n}(r) r^2 \de r= Z,N$. 

By recalling from Eq. (\ref{Ejqrop}) the operator $T_{a1}$ 
which induces the dipole $1^-$ excitations in $\mu^- - e^-$, 
we find that it can be cast in the form 
(\ref{Eomg}) with 
\[ 
f(r) = 6\tilde{g}_Vf_V\sqrt{12\pi}j_1(qr)  
 \] 
and $c_{p}=c_a(1/2)\equiv c_{ap}$, $c_n=c_a(-1/2)\equiv c_{an}$. 
Then, using recursion relations of the Bessel functions, we find that
\begin{equation} 
\frac{1}{r^2}\frac{\de}{\de r} [f(r)r^2] = 6\tilde{g}_Vf_V\sqrt{12\pi} qj_0(qr) . 
\end{equation} 
Notice that the proton, neutron form factor $F_{p,n} (q)= \langle j_0(qr) \rangle_{p,n} $. 
Under the above circumstances Eq.~(\ref{Eetag}) reads
\begin{equation} 
\eta_a = \tilde{g}_Vf_V 4\sqrt{3\pi} q %\sqrt{3/4\pi}q 
\left[ \frac{c_{ap}Z}{A} F_p(q) + \frac{c_{an}N}{A} F_n(q)  \right] 
. 
\end{equation} 
The form factors are calculated numerically using the HF ground-state densities.

The above prescription is a generalization of the method used 
in Ref.~\cite{VaS1981a}, where a purely isoscalar field ($c_p=c_n$) 
was assumed and a specific form
of the function $f(r)$ was utilized, i.e. $f(r)=r^3$. 
In Eq.~(\ref{Eetag}) the values of $c_{p,n}$ and the form of $f(r)$
are arbitrary. However, for the above mentioned isoscalar field, 
the present prescription leads to that given in Refs.~\cite{VaS1981a,ASS03}.  

\begin{figure} 
\includegraphics[width=0.8\columnwidth]{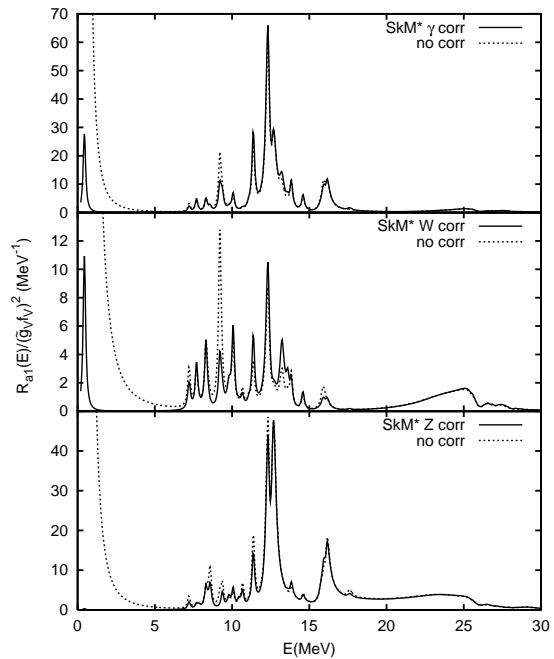} 
\caption{ 
\label{Fdip}  
The dipole distributions $R_{a1}(E)$, for $\gamma$-photon and $W$-boson
exchange diagrams of the $\mu^-\to e^-$ conversion in $^{208}$Pb. The
results have been calculated for dipole operator $T_{a1}$ (dotted line).
In order to estimate the spurious CM contribution of this operator, 
%with the MSk7 force and for $\Gamma = 0.2$~MeV, and for the 
we also show the distribution strength (full line) of the corresponding 
corrected operator given by Eq. (\ref{Eomgc}).} 
\end{figure}

In Fig.~\ref{Fdip}, we plot the dipole distributions $R_{a1}(E)$ of 
Eq.~(\ref{Espr}), for photon, $W$- and $Z$-boson exchange diagrams,
calculated by using the corrected and uncorrected operator 
(they have been obtained with the SkM* force and for 
$\Gamma = 0.2$~MeV). One can see that most of the spurious strength 
below $\approx$6~MeV has been removed. 
The strength distributions above 20~MeV are practically unaffected. 
The strength between 6 and 20~MeV 
appeares somewhat redistributed. The effect of the correction 
appears strongest in the case of the 
$W$-boson exchange mechanism. 
We should note that the radial mesh used in the CRPA calculation, 
with $\Delta r=0.34$~fm, may not be fine enough to yield 
completely converged results in this energy region \cite{HaS2002b}. 
Numerical inaccuracies of such origin may be the reason why strength 
appears to be added at around 13~MeV ($W$-boson case), rather than removed, 
after using corrected operators. 
A result that persists when other interactions are employed, 
%or even when non-self-consistent 
%RPA is used \cite{PapaXX}, 
is that the pygmy dipole state below 10~MeV is strongly affected by the correction 
in the photonic and $W$ cases.

%%%%%%%%%%%%%%%%%%%%%%%%%%%%%%%%%%%%%%%%%%%%%%%%%%%%%%%%%%%%%%%%%
%%%%%%%%%%%%%%%%%%%%%%%%%%%%%%%%%%%%%%%%%%%%%%%%%%%%%%%%%%%%%%%%%

In Table~\ref{table1} we list 
the portion of transition strength removed from  
the total contaminated $1^-$ transition strength $S_{a 1}$ 
($s^{\textrm{sp}}_{\textrm{tot}}$), as well as the portion 
of the strength removed from above 6~MeV excitation energy 
$s^{\textrm{sp}}_{>6{\textrm{MeV}}}$ (with respect to the uncorrected 
strength above 6~MeV),  
for SkM* and for the three 
examined channels. 
For all three mechanisms, about 90\% of the total transition rate 
was spurious. 
We expect this result to be independent of the interaction used, 
because the spurious state at low energy  
always dominates the isoscalar dipole strength distribution 
(which contributes in all three mechanisms) and because the 
corrected operators are, by construction,  
most effective for this state (thus removing practically 
all its strength). 
We were not able to demonstrate this in the particular cases of SGII and MSk7, 
because the energy of the 
spurious state in these cases was found imaginary, within the present calculation. 
In other words, we were not able to evaluate 
and take into account properly the strength of the spurious state, 
before or after the correction. 
  
From Table~\ref{table1} we notice that for $s^{\textrm{sp}}_{>6{\textrm{MeV}}}$ 
and for $a=\gamma$ the result is small in absolute value, but negative. 
It represents the numerical accuracy of our calculation and, being small, 
it indicates that the degree of self consistency reached by our HF+CRPA model 
is sufficient to achieve a satisfactory separation of the spurious 
transition. 
For the $W$ and $Z$ cases, however, spurious admixtures of more than 6\% 
are found above 6~MeV. 
These numbers vary when different Skyrme interactions are used, with their 
values remaining below 10\%. 
(As mentioned before, they are not free of numerical inaccuracies.)  
One should apply the same treatment in the case of other, not self-consistent RPA methods, 
where the lowest $1^-$ state is shifted artificially, 
by means of additional parameters, to zero energy. 
It is possible that larger corrections would be obtained, above 6~MeV. %~\cite{PapaXX}. 
Such a result would mean that excluding the lowest $1^-$ state from the calculation 
of the incoherent rate, as was done on Ref.~\cite{SFK97}, 
would not be an adequate treatment.  
As can be seen from 
Fig.~2 of Ref.~\cite{SFK97}, a considerable amount 
of spurious strength is distributed at higher excitation energies in the case of the QRPA 
calculations reported there. 

%%%%%%%%%%%%%%%%%%%%%%%%%%%%%%%%%%%%%%%%%%%%%%%%%%%%%%%%%%%%%%%%%%%%%%%%%%%

\begin{table} 
\begin{center} 
\begin{ruledtabular} 
\begin{tabular}{l|ccc}   
%\hline 
%\hline 
%& & &    \\
 $\mu - e$ mechanism &  $\gamma$ & $W$  & $Z$      \\ 
\hline 
%& & &   \\
$s^{\textrm{ sp}}_{\textrm{ tot}}(\% )$    &  86.9    & 96.3 & 90.5   \\  
%& & &   \\
$s^{\textrm{ sp}}_{>6{\textrm{ MeV}}}(\% )$&  -1.3    & 7.8  & 6.1     \\  
%\hline 
%\hline 
\end{tabular} 
\end{ruledtabular} 
\end{center} 
\caption{% 
\label{table1} 
Percentage 
of the total $1^-$ transition strength $S_{a 1}$ ($s^{\textrm{ sp}}_{\textrm{ tot}}$) 
and of the strength above 6~MeV ($s^{\textrm{ sp}}_{>6{\textrm{ MeV}}}$)  
consumed by 
spurious transitions, for the interaction SkM*, and for 
the three channels $\gamma$, $W$, $Z$. }  
\end{table} 

%%%%%%%%%%%%%%%%%%%%%%%%%%%%%%%%%%%%%%%%%%%%%%%%%%%%%%%%%%%%%%%%%%%%%%%%%%%

\section{Summary and conclusions} 
\label{Sum} 

In the present paper we have focused on the investigation of the incoherent rate of the
exotic $\mu^- - e^-$ conversion in the heavy nuclear target $^{208}$Pb. We employed,
for the first time in this process, the Continuum-RPA method which is appropriate for
explicit construction of the excited states lying in the continuum spectrum of the
nuclear target. We mainly focused on the distribution of the transition 
strength as a function of the excitation energy of the target for energies up to 30~MeV.

We have investigated in detail the various transition strength distributions 
coming from natural-parity $ph$ excitations up to $L=4$ by using two different 
Skyrme interactions. 
We found that a significant portion of the
incoherent $\mu^- - e^-$ rate comes from high-lying nuclear excitations. 
A similar study could be done for unnatural-parity 
transitions. 

The spurious $1^-$ admixture
was eliminated by 
constructing the purified dipole operators of the $\mu^- - e^-$ conversion 
within the collective model. 
This study provided us with the
interesting result that the greatest portion of the $1^-$ transition strength is due
to the spurious CM excitation, a result in agreement with that of an exact method 
constructed recently \cite{Civi-02} for removing spurious 
contaminations.
The latter is a significant result for the $\mu^- - e^-$ conversion experiments searching
for the coherent rate.

\begin{acknowledgments} 

This work has been supported 
by the Deutsche Forschungsgemeinschaft through contract SFB 634 
  and 
by the IKYDA-02 project% 
.
\end{acknowledgments} 

%%%%%%%%%%%%%%%%%%%%% REFERENCES %%%%%%%%%%%%%%%%%%% 

%%%%%%%%%%%%%%%%%%%%%%%%%%%%%%%%%%%%%%%%%%%%%%%%

\end{document}